\documentclass[%
 aip,
 amsmath,amssymb,
preprint,%
]{revtex4-2}

\usepackage{graphicx}
\usepackage{dcolumn}
\usepackage{bm}

\usepackage{calrsfs}
\DeclareMathAlphabet{\pazocal}{OMS}{zplm}{m}{n}

\usepackage[utf8]{inputenc}
\usepackage[T1]{fontenc}
\usepackage{mathptmx}
\usepackage{etoolbox}

\usepackage{xcolor} 

\definecolor{darkgreen}{rgb}{0.01, 0.6, 0.01}

\newcommand{\revA}{\textcolor{black}} 
\newcommand{\revB}{\textcolor{black}} 
\newcommand{\revC}{\textcolor{black}} 

\makeatletter
\def\@email#1#2{%
 \endgroup
 \patchcmd{\titleblock@produce}
  {\frontmatter@RRAPformat}
  {\frontmatter@RRAPformat{\produce@RRAP{*#1\href{mailto:#2}{#2}}}\frontmatter@RRAPformat}
  {}{}
}%
\makeatother

\begin{document}


\title[Velocity Reconstruction in Puffing Pool Fires with PINNs]{Velocity Reconstruction in Puffing Pool Fires with\\ Physics-Informed Neural Networks}

\author{Michael Philip Sitte}
  \affiliation{Siemens Mobility Austria GmbH, Leberstrasse 34, 1110 Vienna, Austria}
\author{Nguyen Anh Khoa Doan}
  \email{n.a.k.doan@tudelft.nl}
 \affiliation{Faculty of Aerospace Engineering, Delft University of Technology, Kluyverweg 1, 2629 HS Delft, Netherlands}

\date{\today}


\begin{abstract} 
Pool fires are canonical representations of many accidental fires, which can exhibit an unstable unsteady behaviour, known as puffing, which involves a strong coupling between the temperature and velocity fields. Despite their practical relevance to fire research, their experimental study can be limited due to the complexity of measuring relevant quantities in parallel. In this work, we analyse the use of a recent physics-informed machine learning approach, called Hidden Fluid Mechanics (HFM), to reconstruct unmeasured quantities in a puffing pool fire from measured quantities. The HFM framework relies on a Physics-Informed Neural Network (PINN) for this task. A PINN is a neural network that uses both the available data, here the measured quantities, and the physical equations governing the system, here the reacting Navier-Stokes equations, to infer the full fluid dynamic state. This framework is used to infer the velocity field in a puffing pool fire from measurements of density, pressure and temperature. In this work, the dataset used for this test was generated from numerical simulations. It is shown that the PINN is able to reconstruct the velocity field accurately and to infer most features of the velocity field. In addition, it is shown that the reconstruction accuracy is robust with respect to noisy data, and a reduction in the number of measured quantities is explored and discussed. This study opens up the possibility of using PINNs for the reconstruction of unmeasured quantities from measured ones, providing the potential groundwork for their use in experiments for fire research.
\end{abstract}


\maketitle


\section{Introduction}
\label{sec:intro}






In fire research, pool fires are characterised by their stabilisation on a horizontal surface, solid or liquid, where the fuel is supplied through evaporation or pyrolysis, which provides a feedback between heat release, heat transfer and fuel supply. In contrast to (jet) flames, fires are further characterised by a low initial Reynolds number of the fuel stream from evaporation or pyrolysis, a buoyant turbulent diffusion flame, and a non-reacting buoyant plume~\cite{Joulain19982691}. The study of pool fires is relevant from both a theoretical and a practical point of view as they constitute a canonical case for fire research that includes most key physical phenomena essential to the dynamics of natural or accidental fires. Consequently, the study of pool fires is of considerable interest in fire safety research, where they have been commonly used, for instance, as a model fire to study the suppression of fires~\cite{Jenft20141} or as ignition sources to study the spread of fires in rail vehicles~\cite{Lonnermark2012, Shi2020103544}. The case of liquid sodium pool fires is relevant for the safety of sodium cooled nuclear reactors~\cite{Newman1983119}. Lastly, similar behaviour to the puffing of pool fires have also been found to play an important role in the spreading of wildfires \cite{Sibulkin197585, Finney20159833}. 

Pool fires have been studied in depth, starting from some early investigations~\cite{Blinov1957, Hottel196141, Thomas1965983} and several extensive reviews are available on the topic~\cite{Emmons1980223, Joulain19982691, Bishop19982815, Tieszen200167}. Here we focus on an unsteady pulsating motion with a well-defined frequency that pool fires can exhibit called "puffing", which has been shown to be the result of a fluid dynamic rather than a combustion instability~\cite{Hamins1996}. In fact, at the onset of puffing, the pool fire undergoes a bifurcation to a globally unstable puffing state driven by baroclinic and buoyant vorticity generation~\cite{Moreno-Boza2018426}. In a recent work, Moreno-Boza et al.~\cite{Moreno-Boza2018426} performed a global linear stability analysis of 2D axisymmetric pool fires to determine the critical condition for the onset of puffing, which was found to occur when exceeding a critical Rayleigh number. While the above-mentioned onset of puffing occurs in the laminar regime, recent progresses on the modelling of turbulent pool fires include several works using large eddy simulation for predictive modelling \cite{Maragkos201722, Maragkos20193927, Ahmed2021237}.

Despite their importance described above, the experimental study of pool fires has been somewhat limited by the difficulties that arise from the complexity of performing measurements on them, as such techniques are often restricted in either the regions that can be investigated or the quantities that can be measured. This becomes even more challenging for transient fire dynamics, such as puffing fires, as they would require simultaneous diagnostics. In this context, the development of data-analysis and postprocessing techniques that can supplement the missing information by, for example, reconstructing unmeasured quantities or augmenting the resolution of the available data is of great interest to fire research. To achieve this, recent techniques in data science and machine learning have shown a great potential \cite{Barwey2019,Barwey2020,Fukami2020b,Callaham2019,Erichson2020,Carter2021,Raissi2020,Nair2019,Saini2016}. For example, methods based on Proper Orthogonal Decomposition (POD) have been used to reconstruct the velocity field from sparse measurements by finding a data-driven based mapping between the measurable quantities (of, for example, sparse sensors) and the velocity field \cite{Callaham2019} or by using feedforward neural networks to achieve a similar task \cite{Nair2019,Erichson2020}. 
For reacting flows, an extension of POD, called Gappy POD, was developed with some success to infer the velocity in regions where it is not measured \cite{Saini2016}. Further attempts were made at using Convolutional Neural Networks (CNNs) to reconstruct velocity fields from OH-Planar Laser Induced Fluoresence (PLIF) data \cite{Barwey2019, Barwey2020}. Despite their success, the approaches in these works required a database with both the measured quantities and those to be reconstructed, to train the machine learning framework. To circumvent this, a recent framework was proposed by Raissi et al.~\cite{Raissi2020}, called Hidden Fluid Mechanics (HFM), where a feedforward neural network is trained with both the governing equations of the system and the measured quantities to perform such a reconstruction task without requiring data of the quantities to reconstruct. \revC{Such a network was coined a "Physics-Informed Neural Network" (PINN) and was originally designed to solve direct PDE problems such as the Navier-Stokes equation \cite{Raissi2019b} or its Reynolds-averaged version \cite{Eivazi2021etmm}.} In the context of experimental measurements reconstruction, as a proof of concept, Raissi et al.\cite{Raissi2020} applied PINN to the K{\'a}rm{\'a}n vortex street in the wake of a cylinder. This problem presented itself as the prediction of a periodic laminar flow, i.e. velocity and pressure fields, from a conserved scalar variable. The so-called Physics-Informed Neural Network was shown able to reconstruct the velocity field in this canonical non-reacting flows. Cai et al.~\cite{Cai2021A102} have applied HFM to a buoyancy-driven flow using experimental data as input. They inferred the velocity and pressure fields above an espresso cup from snapshots of the temperature field recorded with tomographic background-oriented Schlieren imaging. However, the result were only validated qualitatively since velocity measurements were not carried out simultaneously with the temperature measurements. 
\revC{A similar PINN-based framework was also applied to the problem of super-resolution, i.e. inferring higher resolution flow features from low resolution ones in space and time\cite{Eivazi2022,Wang2022017116}. It was shown that the PINN could reconstruct the higher resolution flow characteristics in space and time for a series of canonical non-reacting flows both from simulation data \cite{Eivazi2022} or from experimental data \cite{Wang2022017116}.} 
Therefore, the HFM approach has only been applied to flows of constant density or small density variations that can be described by a Boussinesq approximation. The applicability of this approach has not been fully demonstrated for reacting flows with large variations and steep gradients of density and temperature as well as significant variations of composition. 

In this work, we propose to extend this HFM methodology of Raissi et al.~\cite{Raissi2020} to the reconstruction of the velocity field in a puffing pool fire, thereby demonstrating the potential of this method in fire research, which deals with complex reacting flows that include chemical reactions, large density variations and buoyancy. This work opens up the use of HFM to practical fire-related applications where obtaining a full set of measurements is extremely challenging. In this first attempt to apply HFM to a reacting flow proposed here, the dataset of measured quantities will be obtained from numerical simulations as this provides all quantities of interest from which only a subset will be used for the development of the HFM framework and as it allows for a thorough cross comparison between the reconstruction from the HFM framework and the values obtained from the simulations. Nevertheless, these simulations are set up as numerical representations of real flames studied experimentally and will be shown to reproduce their behaviour accurately \cite{Moreno-Boza2018426}. The reconstruction problem in scope is the reconstruction of the full velocity field from measurements of density, pressure and temperature. 

The remainder of this paper is organised as follows. Section \ref{sec:simu} presents the set-up of the numerical simulations of the pool fire. Section \ref{sec:HFM} provides details on the HFM framework. Numerical results of the pool fire simulations are provided in Section \ref{sec:res_CFD} and the velocity reconstruction with HFM is discussed in Section \ref{sec:res_recons}. Additional comments on the robustness of the HFM framework with respect to noisy data and a reduced number of measured quantities are also discussed. The final section summarises the results obtained and provides directions for future work.

\section{Methods} 
\subsection{Computational fluid dynamics}
\label{sec:simu}
Simulations of the pool fire were carried out using the computational fluid dynamics (CFD) toolbox OpenFOAM-7 with the solver \emph{fireFoam}~\cite{openfoam7}. The governing equations are the continuity equation, Navier-Stokes equations, species mass fraction equations (for each chemical specie $k$) and enthalpy equation \cite{PoinsotVeynante2005}:
\begin{equation}
\frac{\partial \rho}{\partial t} + \frac{\partial \rho u_i}{\partial x_i} = 0
\label{eq:continuityEq}
\end{equation}
\begin{equation}
\frac{\partial \rho u_i}{\partial t} + \frac{\partial \rho u_i u_j}{\partial x_j} 
= 
- \frac{\partial p}{\partial x_i} + \frac{\partial \tau_{ij}}{\partial x_j}+ \rho g_i
\label{eq:NavierStokesEq}
\end{equation}
\begin{equation}
\frac{\partial \rho Y_k}{\partial t} + \frac{\partial \rho u_i Y_k}{\partial x_i} 
= 
\frac{\partial }{\partial x_i} \left[ \rho D \frac{\partial Y_k }{\partial x_i} \right] + \rho\dot{\omega}_k
\label{eq:YEq}
\end{equation}
\begin{equation}
\frac{\partial \rho h}{\partial t} + \frac{\partial \rho u_i h}{\partial x_i} 
= 
\frac{\partial }{\partial x_i} \left[\rho \alpha \frac{\partial h }{\partial x_i} \right]
\label{eq:hEq}
\end{equation}
where $\tau_{ij}$ is the molecular stress tensor, $\dot{\omega}_k$ is the chemical reaction rate, $D$ is the molecular diffusivity and $\alpha=\lambda/(\rho C_p)$ is the thermal diffusivity. 

\begin{figure}[!ht]
    \centering
    \includegraphics[trim={0 0 0 0},clip, width=35mm]{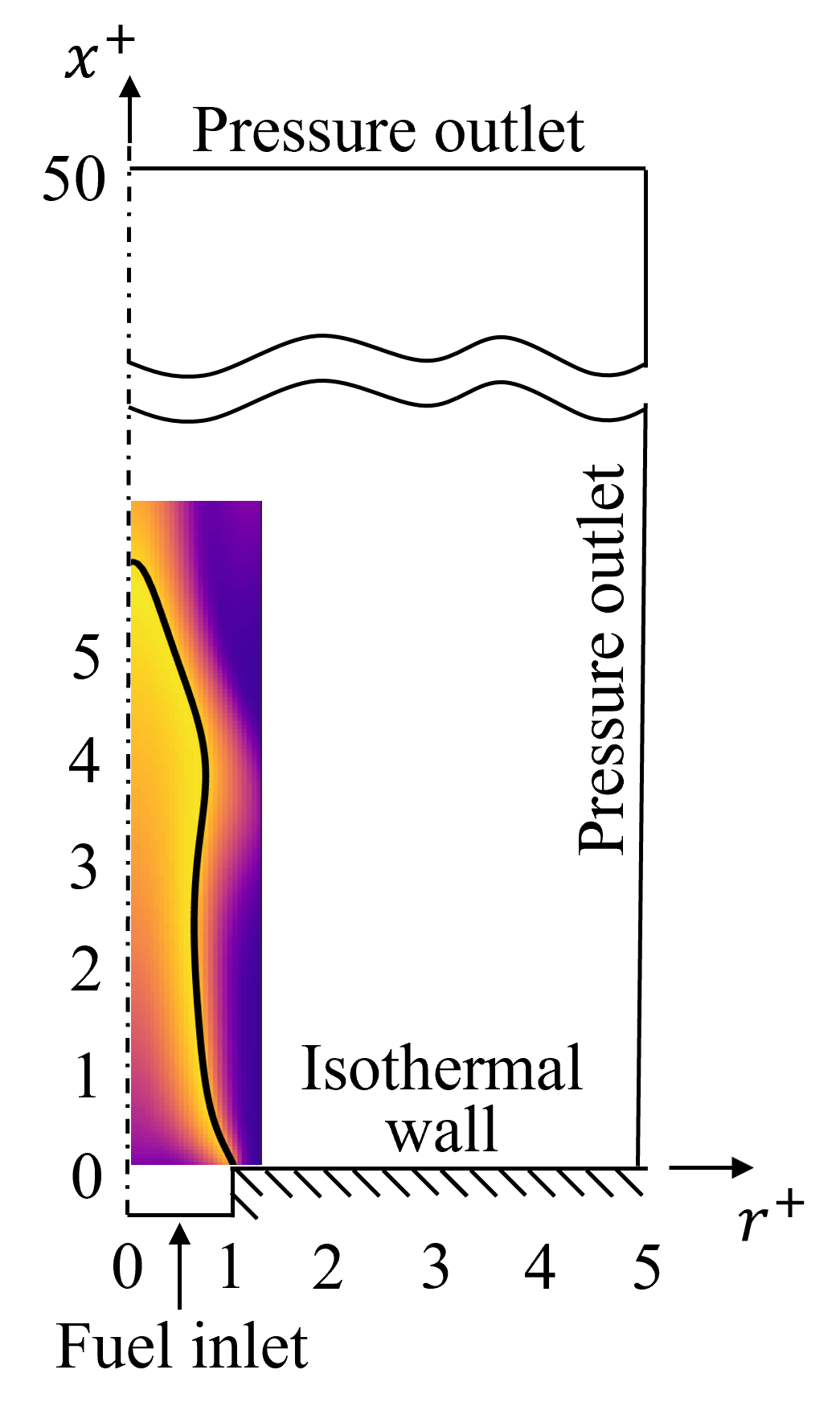}
    \caption{Schematic of the numerical domain. Coordinates normalised by pool radius $a$.}
    \label{fig:numerical_domain}
\end{figure}

A 2D axisymmetric pool fire with n-heptane fuel ($Y_\mathrm{F}=Y_\mathrm{C7H16}$) and at ambient conditions ($p_0= 100~000~\mathrm{Pa}$, $T_0= 300~\mathrm{K}$, $Y_\mathrm{O2}= 0.233$, $Y_\mathrm{N2}= 0.767$) was investigated, equivalent to the n-heptane flames with an isothermal brass base plate studied experimentally and numerically by Moreno-Boza et al.~\cite{Moreno-Boza2018426}. A schematic of the numerical domain is shown in Fig.~\ref{fig:numerical_domain}. The domain size was $50a$ in axial and $5a$ in radial direction, where $a$ is the fuel pool radius. A 2D structured grid was used with $600 \times 130$ grid points. \revB{In radial direction the pool radius $a$ was resolved by 91 grid points. In the region of interest, where the flame was located, the same resolution was used in axial direction. So, depending on the pool diameter, the grid resolution was in the range from approximately 0.088 to $0.11~\mathrm{mm}$}, which is sufficient to resolve all scales of the flow and the reaction zone. \revB{This resolution was the result of a grid sensitivity study, where the resolution was refined until key characteristics of the flame, i.e. stable flame length and puffing frequency in the transient case}, did not change any more and grid convergence was achieved. Numerical schemes were of second order in space and first order in time with a fixed time step of $10^{-5}~\mathrm{s}$, so that $\mathrm{CFL}<0.1$, and using a fractional step scheme \revB{with adaptive step size for the chemical source term. This was found to be sufficiently accurate to replicate the characteristics of the flames observed in the experiments and effective to avoid numerical instabilities.}

In the present modelling approach, radiation was \revB{not explicitly included as a sink term in the enthalpy equation. Instead radiative heat loss was accounted for by reducing the heat of combustion of the chemical reaction. Turbulence modelling did not need to be considered, since the flame is} in the laminar regime at the onset of puffing. Combustion chemistry was modelled as the irreversible single-step reaction,
\begin{equation}
\mathrm{C_7 H_{16} + 11~ O_2 \rightarrow 7~ CO_2 + 8~ H_2 O}
\label{eq:chemistry}
\end{equation}
whose rate constant is given by the Arrhenius law,
\begin{equation}
K = B T^\beta \exp(-T_A/T)
\end{equation}
with the model constants $\beta = 0$, $T_A = 12~000~\mathrm{K}$ and $B= 5.5\times10^7~\mathrm{m^3/(mol\cdot s)}$. The choice of $T_A$ and $B$ was made to best fit the laminar flame speed curve for heptane-air mixture with the single-step mechanism, following the reasoning of Fernandez-Tarrazo et al.~\cite{Fernandez-Tarrazo200632}, but without correcting heat release rate (HRR) and $T_A$ with equivalence ratio $\phi$. This modelling of the chemical reaction was chosen for the sake of simplicity and it is sufficient for the present case of a laminar diffusion flame characterised by high Damköhler number and controlled by mixing. Indeed, Moreno-Boza et al.~\cite{Moreno-Boza2018426} reproduced the puffing behaviour correctly using infinitely fast chemistry. Reproduction of the laminar flame speed $S_L(\phi)$ and the $\mathrm{HRR}(\phi)$ over a broad range of equivalence ratios, as suggested by Fernandez-Tarrazo et al.~\cite{Fernandez-Tarrazo200632}, is less important for a diffusion flame where the reaction occurs at stoichiometric conditions.

Density is given by the ideal gas law. Viscosity is computed from Sutherland's law, $\mu = A_s \sqrt{T} / (1 + T_s/T)$, independent of species composition with $A_s= 1.672 \times 10 ^{-6}~\mathrm{kg/(m\cdot s\cdot K^{1/2})}$ and $T_s= 170.67~\mathrm{K}$. Molecular and thermal diffusivity are computed based on the assumptions of unity Lewis number, $\mathrm{Le}=\alpha/D=1$, and constant Prandtl number, $\mathrm{Pr}= \mu/ (\rho \alpha) = 0.7$. Specific heat, $C_p$, and enthalpy, $h$, are obtained from NASA polynomials but modifying the formation enthalpy of n-heptane to reduce the flame temperature to approximately 1750~K to account for the effect of radiative heat loss on the flame, following previous work~\cite{Moreno-Boza2018426}.

The liquid surface is modelled as a boundary condition following Moreno-Boza et al.~\cite{Moreno-Boza2018426}. The liquid surface is assumed to be at the boiling temperature of n-heptane, $T_B = 371.5~\mathrm{K}$. Walls are assumed isothermal at the ambient temperature $T_0$. The fuel mass flow rate at the liquid surface is determined by the evaporation rate relating the conductive heat flux to the liquid with the surface normal velocity $u_n$:
\begin{equation}
\rho u_n = \frac{1}{l_V} \lambda \frac{\partial T}{\partial n}
\end{equation}
where $l_V = L_V + C_L (T_B - T_0)$ with the latent heat $L_V = 360~000~\mathrm{J/kg}$ and the liquid heat capacity $C_L = 2~240~\mathrm{J/(kg\cdot K)}$. The jump condition at the phase interface for fuel mass fraction (species index $k = \mathrm{F}$) is
\begin{equation}
\rho u_n 
= \rho u_n Y_\mathrm{F} 
- \rho D \frac{\partial Y_\mathrm{F}}{\partial n}
\end{equation}
and for all other species ($k \neq \mathrm{F}$),
\begin{equation}
0
= \rho u_n Y_k 
- \rho D \frac{\partial Y_k}{\partial n}
\end{equation}
In the experiments, the fuel surface was maintained $0.75 \pm 0.25~\mathrm{mm}$ below the surface of the brass base plate to avoid fuel spillage and the flame behaviour was found to be very sensitive to that value \cite{Moreno-Boza2018426}. Therefore, this vertical offset between the isothermal base plate and the liquid surface is included in the present simulations.

\subsection{Reconstruction method}
\label{sec:HFM}

\begin{figure*}[!ht]
    \centering
    \includegraphics[width=340pt]{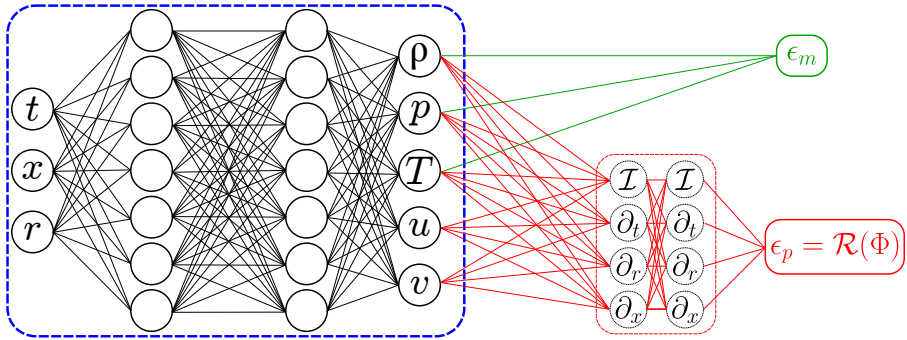}
    \caption{Schematic of the HFM framework. \revA{The dashed blue box indicates the actual neural network. The dashed red box indicates the differential operators  ($\pazocal{I}$ indicates the identity operation, $\partial_{\cdot}$ is the partial derivative), obtained via automatic differentiation, required to compute the physical residual, $\epsilon_p$. The green box indicates the calculation of the measurement reconstruction error, $\epsilon_m$}.}
    \label{fig:HFM_schema}
\end{figure*}

To reconstruct the unmeasured quantity, the approach called Hidden Fluid Mechanics proposed by Raissi et al.~\cite{Raissi2020} will be used. This approach relies on a Physics-Informed Neural Network  \cite{Raissi2019a}, illustrated in Fig.~\ref{fig:HFM_schema}, and uses a neural network to infer the measured and unmeasured quantities in a given flow, which is governed by the (reacting) Navier-Stokes equations, listed in Sec. \ref{sec:simu}. To achieve the inference of unmeasured quantities, the PINN relies on two sources of information: (i) the data which is related to the \textit{measurable} quantities and (ii) the governing equations of the flow, described in Sec.~\ref{sec:simu}. A PINN is a conventional feedforward neural network (blue box in Fig.~\ref{fig:HFM_schema}) which is trained with a specific loss function that accounts for the governing physical equations, \revB{which we will call the "physical residual error" $\epsilon_p$ (red part in Fig.~\ref{fig:HFM_schema}) and the measurement reconstruction error ($\epsilon_m$, green part)}. Feedforward neural networks map the input to the output, as shown in Fig.~\ref{fig:HFM_schema} where an input, two hidden and an output layer are represented. The network is termed feedforward as the output of a given layer is not fed back into the input or preceding layers. Each hidden layer consists of neurons which are fully connected, meaning that each neuron in a given layer $l-1$ is connected to all neurons in the following layer $l$ through a weight matrix $\bm{W}_l$. Therefore, the intermediate output of the hidden layer $l$ can be written as $\bm{Z}_l = \bm{W}_l^T \bm{X}_{l-1} + \bm{b}_l$ where $\bm{X}_{l-1}$ is the output of the layer $l-1$ and $\bm{b}_l$ is the bias in layer $l$. Following this, nonlinearities are introduced through the element-wise activation function $g$, so that $\bm{X}_l = g(\bm{Z}_l)$. For what follows, the activation function used for all hidden layers will be the swish function \cite{Ramachandran2017} with a linear activation in the final output layer. \revB{The choice of the swish activation function was made following the propositions on HFM architecture in previous work~\cite{Raissi2020}. The swish function was found to provide sufficient accuracy for the present cases as well, which will be discussed in Sec.~\ref{sec:res_recons}. Other activation functions, such as the sigmoid and ReLU activation functions, were also tested but in those cases the accuracy of the resulting PINN was smaller, in line with previous findings \cite{Ramachandran2017}. The interested reader is referred to this latter reference for details on how the swish activation function outperforms previous activation function.}

\begin{figure*}[!ht]
    \centering
    \includegraphics[width=350pt]{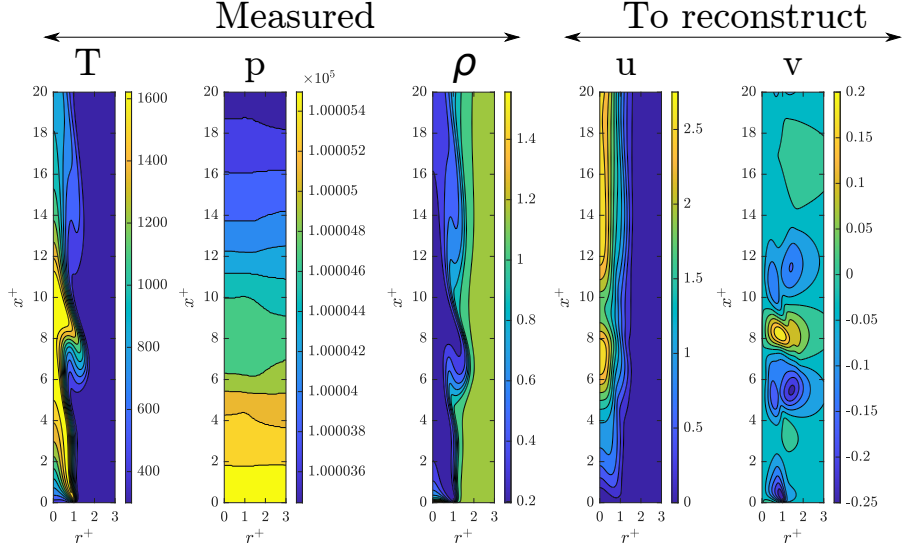}
    \caption{Typical fields that are used for the reconstruction problem. $T$, $p$, $\rho$ are considered as measurements while $u$ and $v$ are quantities to be reconstructed.}
    \label{fig:HFM_measurements}
\end{figure*}

In the HFM architecture, the PINN takes as input a space-time location $(\bm{x},t)$ (here, in axisymmetric 2D, $\bm{x}=(x,r)$) and outputs the flow state vector at that location, i.e. its output is $\widetilde{\Phi}=[\rho,p ,T ,\bm{u}, \bm{Y}]$, where $~\widetilde{\cdot}~$ indicates a prediction from the PINN, $\bm{u}=(u,v)$ is the velocity vector, with axial component $u$ and radial component $v$, and $\bm{Y}$ is the chemical composition vector. For simplicity, in this first attempt at using the HFM for reacting flows, the species mass fraction will not be considered in the reconstruction problem and, therefore, the considered flow state is just $\widetilde{\Phi}=[\rho,p ,T ,u, v]$.
In the specific reconstruction problem considered here, it should be stressed that, for the training of the PINN, only a subset of these outputs have associated \textit{target data} (i.e. the \textit{measured} quantities). Such measured states will subsequently be noted $\bm{\phi}_m$ for the target data and $\widetilde{\bm{\phi}}_m$ for the PINN prediction, with their associated space-time locations noted as $(\bm{x_m},t_m)$. In what follows, it will be considered that the measurable quantities are $\bm{\phi}_m=[\rho, p,T]$ unless mentioned otherwise. \revC{A visual representation of the quantities that are considered measured and the ones to be reconstructed is provided in Fig. \ref{fig:HFM_measurements}.}

To enable the PINN to reconstruct the other \textit{unmeasured} quantities (here the velocity field $\bm{u}$), the loss function used to train the PINN includes the residual of the reacting Navier-Stokes equations, noted $\pazocal{R}$, which is estimated at collocation points (noted $(\bm{x_c},t_c)$) spread over the time-space domain covered by the simulation. \revA{It should be noted that given the specific choice made here in terms of the quantities taken into account (species not considered and $T$ being measured), it becomes possible to only consider a subset of the complete reacting Navier-Stokes equations, namely, the continuity and momentum equations. Indeed, if $T$ is measured, these equations would form a closed set of equations (for the density, pressure and the velocity fields) and, therefore, the species transport equations and the enthalpy equation are not considered in what follows.}
Thus, the loss function to train the PINN is:
\begin{align}
\pazocal{L}= \underbrace{\frac{1}{N_m}\sum_{n=1}^{N_m} |\bm{\phi}_m(\bm{x}_m^n,t_m^n) - \bm{\tilde{\phi}}_m(\bm{x}_m^n,t_m^n) |^2}_{\epsilon_m} \nonumber \\
+ \underbrace{\frac{1}{N_c}\sum_{i=1}^{N_c} |\pazocal{R}(\widetilde{\bm{\Phi}}(\bm{x}_c^i,t_c^i)|^2}_{\epsilon_p}
\label{eq:loss}
\end{align}
In the equation above, the first term, \revB{the measurements reconstruction error} $\epsilon_m$ (green part in Fig. \ref{fig:HFM_schema}), corresponds to a standard mean-squared-error between the target data \revB{(the measured fields)} and the prediction of the PINN for the $N_m$ space-time locations $\left\{(\bm{x}_m^n,t_m^n)\right\}$ where \revB{those measurements data} is available. It should again be emphasised that only a subset of all the flow states are considered measurable and therefore $\bm{\phi}_m$ does not account for the full flow state. The second term ($\epsilon_p$, red part in Fig. \ref{fig:HFM_schema}) represents a physics-based loss which is the residual of the (reacting) Navier-Stokes equations computed using the prediction of the PINN at arbitrary space-time collocation points $\left\{(\bm{x_c^i},t_c^i)\right\}$. This second term enables the network to identify suitable predictions for the unmeasured quantities that satisfy the governing equations. \revC{The role of the collocation points is, therefore, quite important as these are the locations where the PINN will try to minimise the residual of the governing equations.} This physical residual is computed using automatic differentiation \cite{Paszke2017} as in past work on PINN applications\cite{Raissi2019b,Raissi2020}. Automatic differentiation enables the efficient computation of the (partial) derivatives of the output (the flow states) with respect to the input (space and time). This allows to estimate all the required partial derivatives needed in the (reacting) Navier-Stokes equations (Eqs~ \eqref{eq:continuityEq}-\eqref{eq:hEq})), therefore, enabling an estimation of the residual of those equations at the collocation locations. \revA{It should be noted here again that, given the specific choice of measured quantities and quantities to be reconstructed, it is only necessary to consider the continuity and momentum equations in the physical residual error calculations and, therefore, only Eqs~ \eqref{eq:continuityEq}-\eqref{eq:NavierStokesEq} are actually considered in the calculation of $\epsilon_p$ in this work.}
Finally, for simplicity, the collocation points are taken to be the same as the space-time locations where target data is available, i.e. $N_c=N_m$ and $\left\{(\bm{x}_c^i,t_c^i)\right\} = \left\{(\bm{x}_m^i,t_m^i)\right\}$.

\revA{Regarding the boundary conditions, previous work on HFM \cite{Raissi2020} showed that an arbitrary domain could be considered for reconstruction. So, there was no need for an explicit knowledge of the boundary condition and the measurement data (from the scalar) was  sufficient for the reconstruction of the unmeasured quantities by the PINN. Similarly, in our work, we did not need to explicitly provide the boundary conditions to the PINN -- the measured quantities were sufficient for the PINN to infer the unmeasured quantities. In general, other problems may exist where the boundary conditions would have to be explicitly specified (for example when trying to solve the direct PDE problem) to ensure the convergence of the PINN towards an appropriate solution \cite{Raissi2019b,Eivazi2021etmm}.}

\begin{figure}[!ht]
    \centering
    \includegraphics[width=67.7mm]{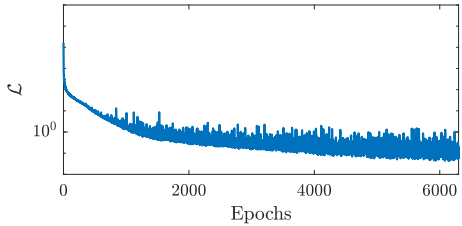}
    \caption{Typical evolution of the loss function (Eq. \eqref{eq:loss}) during the training process.}
    \label{fig:loss_evol}
\end{figure}

In the present work, following Raissi et al.\cite{Raissi2020}, \revB{a feedforward neural network of 20 hidden layers with 300 neurons each is used. The swish activation function is used for all neurons in all layers except in the last layer where the linear activation function is used.} The training is performed in two stages: (i) the network is pre-trained using only the available data, i.e., the loss function only contains the first term in Eq.~\eqref{eq:loss}, which allows for a rapid partial weight optimisation as it corresponds to a traditional supervised training process with a mean-squared error (MSE) loss; (ii) all the network weights are optimised using the full loss function, as in Eq.~\eqref{eq:loss} enabling the reconstruction of the unmeasured quantities. All the training processes are performed with the Adaptive Moment Estimation (ADAM) optimiser \cite{Kingma2015} using a learning rate of 0.001 with a batch size of 10~000. During this training, the value of the loss function is monitored and the training is stopped when it has reached a plateau indicating a fully trained network. A typical evolution of the loss function is shown in Fig. \ref{fig:loss_evol} where the training process has converged after approximately 6000 epochs.

\section{Results} 
\label{sec:res}
\subsection{Flame Dynamics from CFD}
\label{sec:res_CFD}

\begin{figure}[!ht]
    \centering
    \includegraphics[trim={0  0 0 0},clip, width=67.7mm]{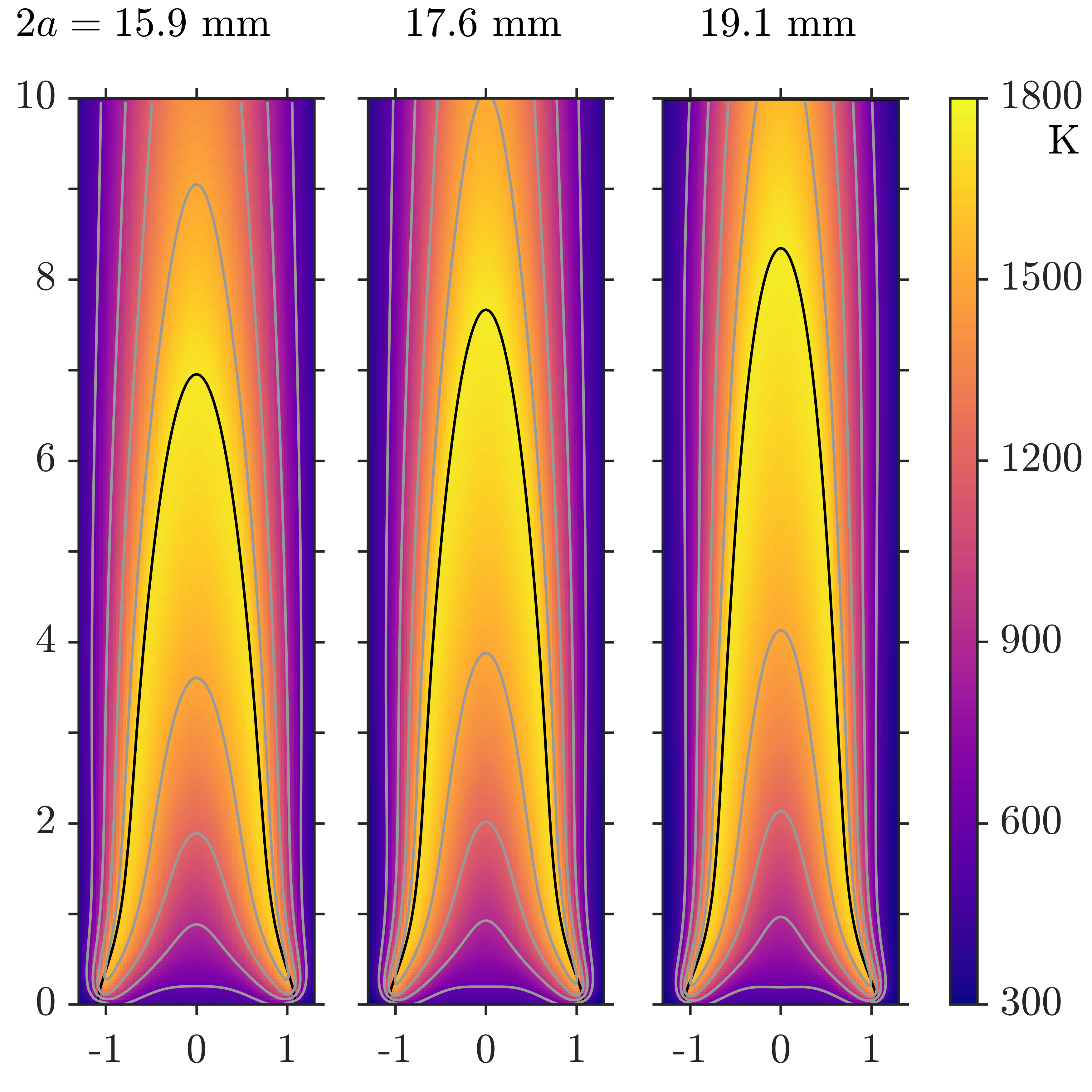}
    \caption{Subcritical flames for different diameters, $2a=15.9$, 17.6 and 19.1~mm. Black line: position of the reaction zone, indicated by the stoichiometric mixture fraction; grey lines: isocontours for $T= 600$, 900, 1200, 1500~K. }
    \label{fig:stable_flames}
\end{figure}

\begin{figure}[!ht]
    \centering
    \includegraphics[trim={0  0  0  0},clip, width=67.7mm]{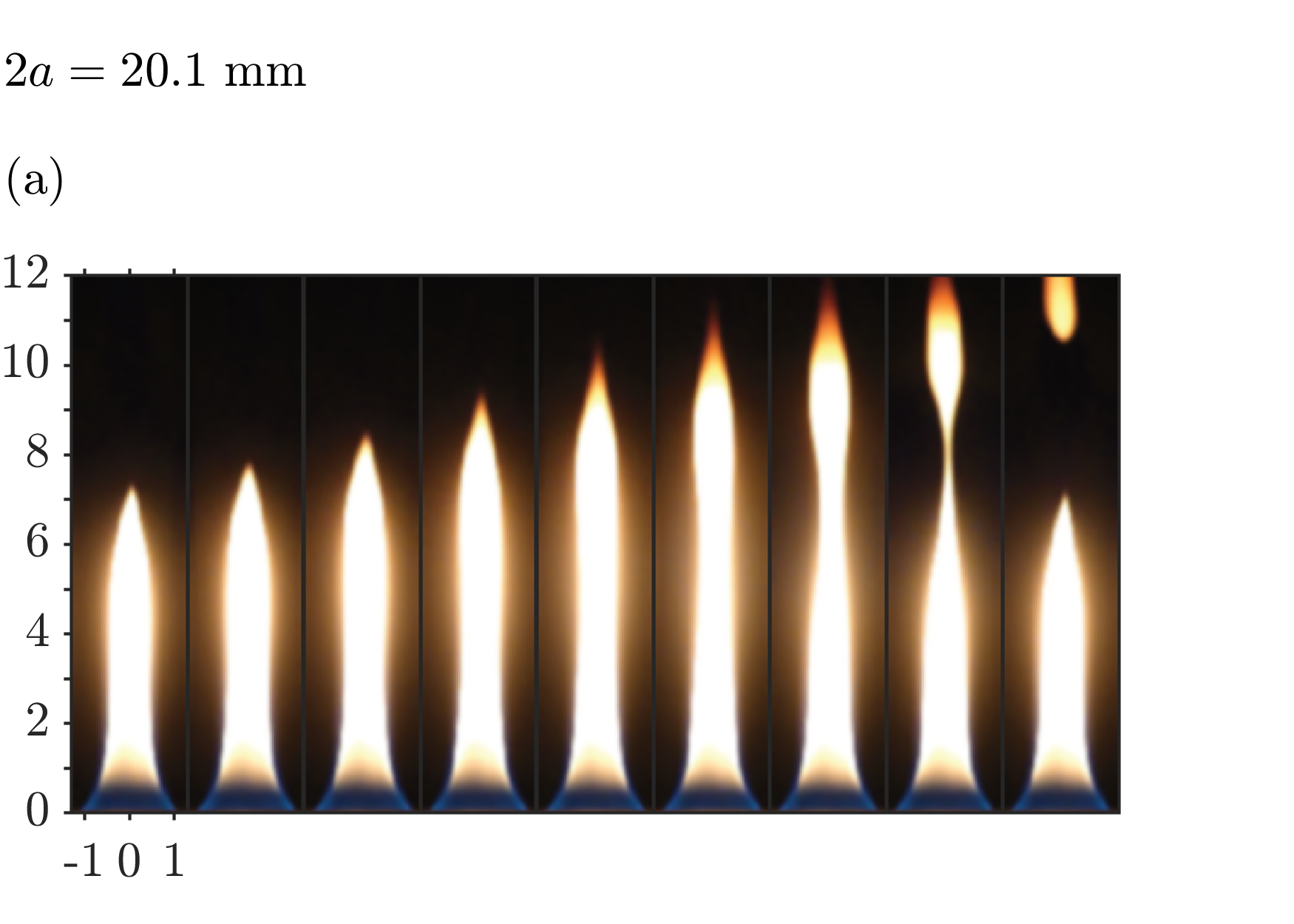} \\
    \includegraphics[trim={0  0  0  15mm},clip, width=67.7mm]{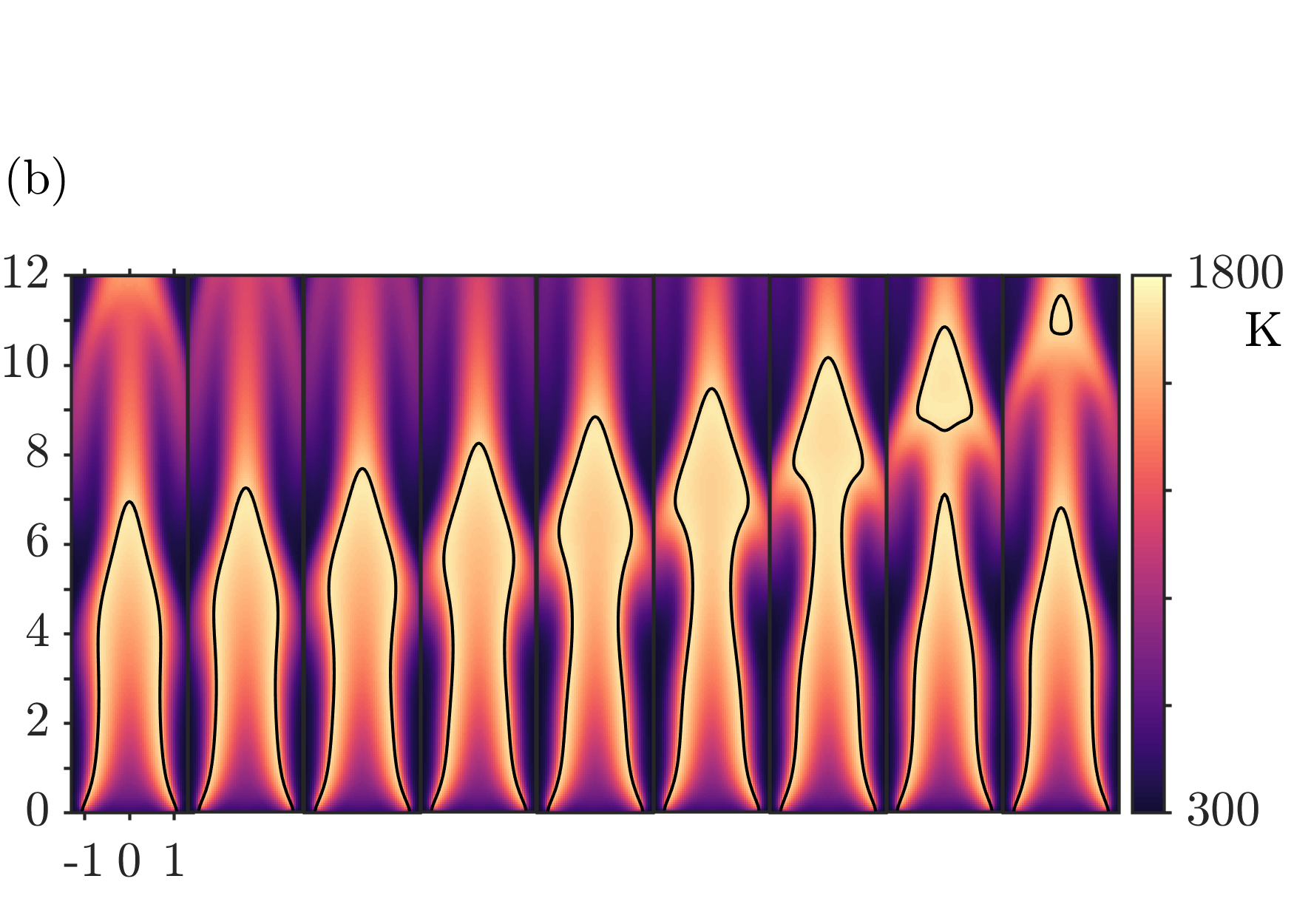} \\
    \includegraphics[trim={0  0  0  0},clip, width=67.7mm]{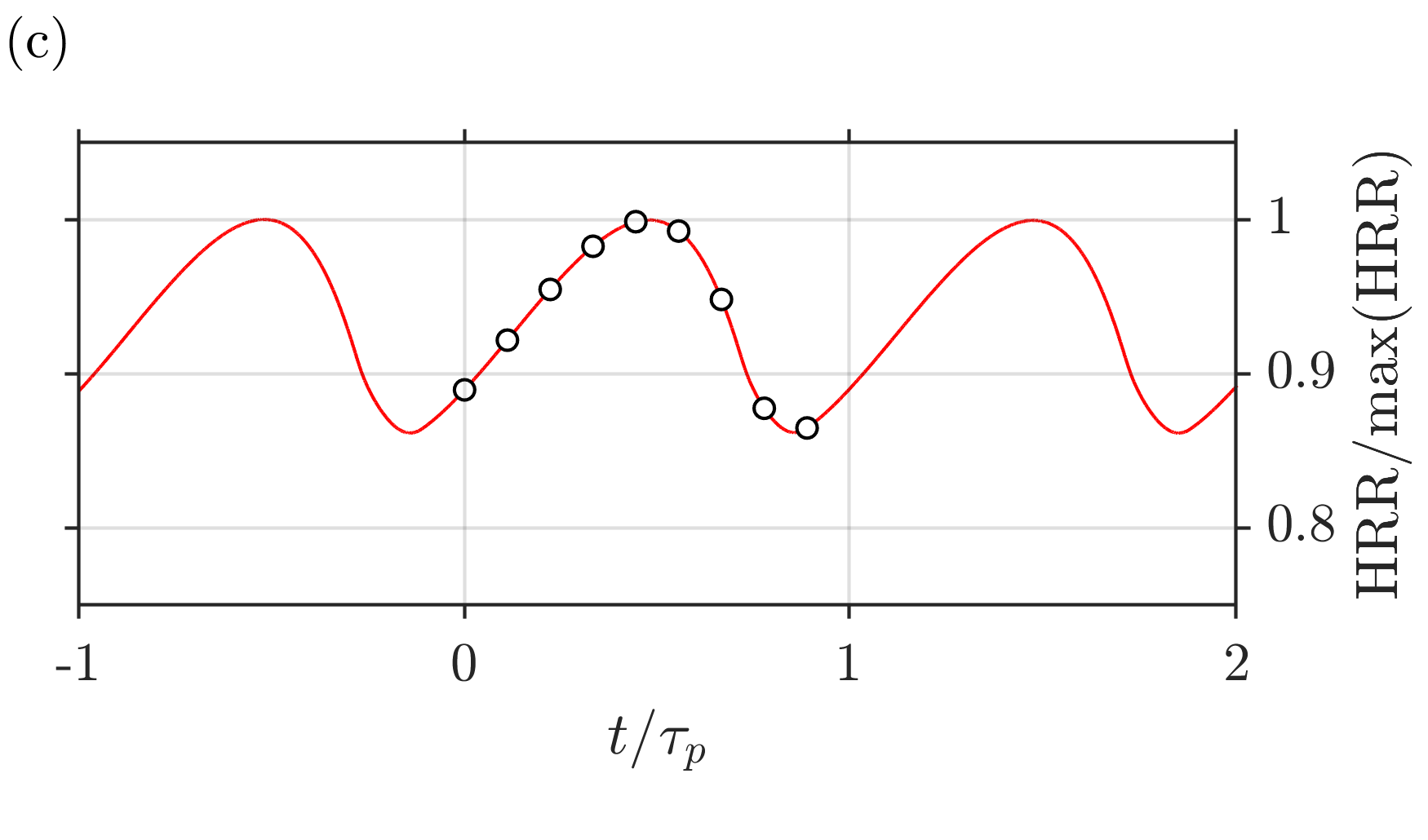}
    \caption{\revA{Puffing cycle of the flame with diameter $2a=20.1~\mathrm{mm}$: (a) photo sequence from the experiments by Moreno-Boza et al.~\cite{Moreno-Boza2018426}, (b) time sequence from the present CFD simulations and (c) the corresponding time evolution of HRR, normalised by period $\tau_p$ and $\max(\mathrm{HRR})$; instances of snapshots are indicated as $\circ$; snapshots are spaced by $\tau_p/9$ to match the spacing of the photo sequence.}}
    \label{fig:puffing_flame}
\end{figure}

The CFD simulations were validated against the experimental and numerical results for the n-heptane flame with isothermal walls from Moreno-Boza et al.~\cite{Moreno-Boza2018426}, notably the reproduction of the flame length, the critical point and the puffing frequency. Figure~\ref{fig:stable_flames} shows temperature contours for the subcritical flames of diameters $2a=15.9$, 17.6 and 19.1~mm with corresponding flame lengths of $7.0a$, $7.7a$ and $8.4a$. The flame length was determined based on the downstream tip of the stoichiometric mixture fraction isocontour (Bilger's definition of the mixture fraction~\cite{Bilger1976155} is used; the stoichiometric mixture fraction is $\xi_\mathrm{st}\approx 0.0622$). These values are in good agreement with the experiments where the flame lengths ranged from $6.7a$ to $8.0a$. In the present simulations, the critical point -- defined as the characteristic diameter where puffing first occurs -- is found at $2a = 20.1$~mm, in line with experimental findings for the isothermal base plate. At this condition, the puffing establishes as a periodic process with a constant frequency of approximately 12.0~Hz, very close to the experimentally determined value of 12.8~Hz. \revA{Figure~\ref{fig:puffing_flame} shows one cycle of this puffing behaviour; (a) a photo sequence of the puffing flame from the experiments by Moreno-Boza et al.~\cite{Moreno-Boza2018426} is compared to (b) a time sequence from the present CFD simulations; additionally, (c) the temporal variation of HRR during the puffing cycle are shown. In this sequence, we can see the dynamics of the puffing flame, where a "cusp" is formed due to buoyancy-driven vorticity generation. Even though photographic images and temperature fields are not fully equivalent,} the numerical results closely resembles the time sequence of the flame recorded in the experiments. 
Therefore, the present simulations have been shown to accurately reproduce the behaviour of the n-heptane pool fire in the vicinity of the critical point, studied by Moreno-Boza et al.~\cite{Moreno-Boza2018426}. 

Excellent agreement of flame length and puffing behaviour suggest that the present highly resolved numerical simulations are a close representation of a real flame. Therefore, the comparison with experimental work is implicit in the present work. For the reconstruction problem discussed in the next section, the puffing flame with $2a= 20.1~\mathrm{mm}$ is considered. The dataset used to train the PINN consisted of 400 snapshots recorded at 1~000 frames per second, corresponding to approximately 5 puffing periods with about 83 snapshots per period.

\subsection{Velocity reconstruction}
\label{sec:res_recons}
In this section, we demonstrate the ability of the PINN to reconstruct the velocity field of the puffing flame from measurements of temperature, density and pressure. The PINN was trained as discussed in Sec.~\ref{sec:HFM} using the dataset obtained from the simulations (and shown in Sec.~\ref{sec:res_CFD}). It should be emphasised that, in the training dataset, only the temperature, density and pressure fields are provided to the network and that it never receives any velocity information. The network then infers both components of the velocity fields using the residual of the reacting Navier-Stokes equations. 

\begin{figure}
    \centering
    \includegraphics[width=67.7mm]{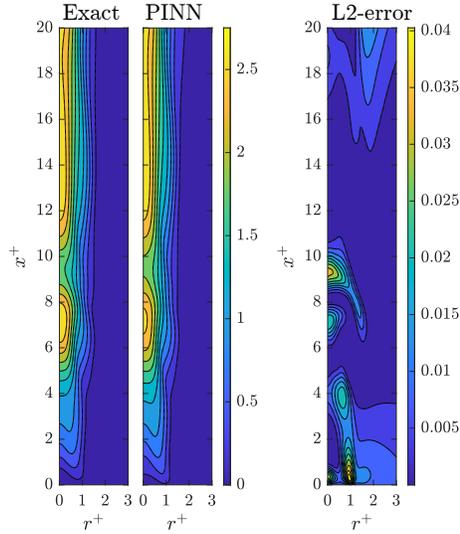}
    \caption{Comparison between actual and reconstructed $u$ velocity for the mid-time snapshot ($t=0.2~\mathrm{s}$). Superscript $+$ indicates coordinates normalised by $a$.}
    \label{fig:recons_velocity_u}
\end{figure}

\begin{figure}
    \centering
    \includegraphics[width=67.7mm]{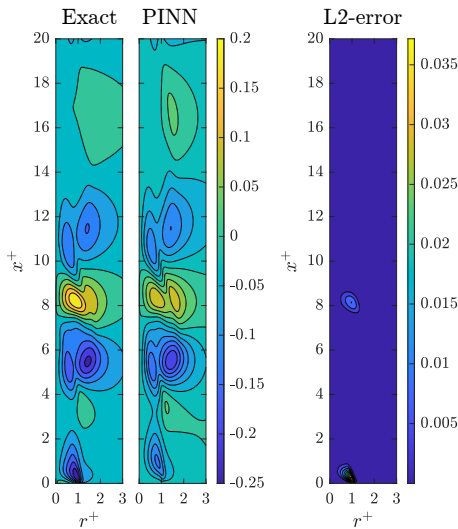}
    \caption{Comparison between actual and reconstructed $v$ velocity for the mid-time snapshot ($t=0.2~\mathrm{s}$).}
    \label{fig:recons_velocity_v}
\end{figure}

\begin{figure}
    \centering
    \includegraphics[width=67.7mm]{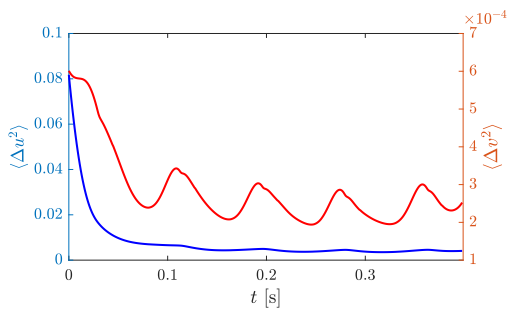}
    \caption{Evolution of MSE of the reconstruction of $u$ and $v$.}
    \label{fig:error_vel}
\end{figure}

The reconstructed velocity field for a snapshot at time $t=0.2~\mathrm{s}$ (which is the mid-time of the database) is shown in Figs~\ref{fig:recons_velocity_u} and \ref{fig:recons_velocity_v} for $u$ and $v$ respectively. As can be seen in those figures, the two components of the velocity field are accurately reconstructed by the PINN and only minor differences can be observed at the base of the flame where strong gradients are present. This shows that the PINN manages to infer the dynamics of the buoyant plume from the residual of the Navier-Stokes equations \textit{without} any observation of the velocity field.
Additionally, the overall $L^2$-error remains small over the majority of the domain and most features of the velocity fields are accurately reconstructed. While this is shown for a specific time instant, the reconstruction accuracy was similar for most snapshots and the mean squared error (averaged over the computational domain) in function of time is shown in Fig.~\ref{fig:error_vel}. It can be seen that the error is overall low with a higher value at the starting time ($t=0~\mathrm{s}$).

\begin{figure}
    \centering
    \includegraphics[width=67.7mm]{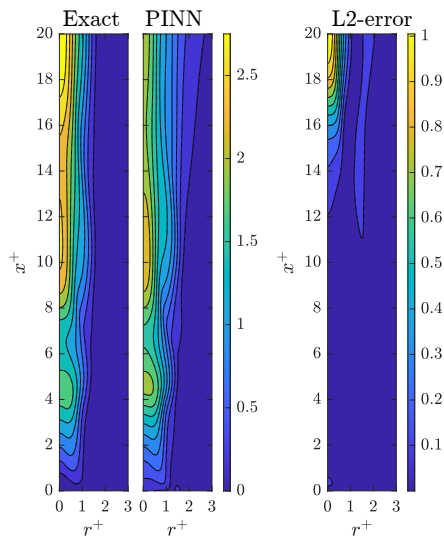}
    \caption{Comparison between the exact and reconstructed $u$ for the snapshot with maximal error ($t=0~\mathrm{s}$).}
    \label{fig:max_err_u}
\end{figure}

\begin{figure}
    \centering
    \includegraphics[width=67.7mm]{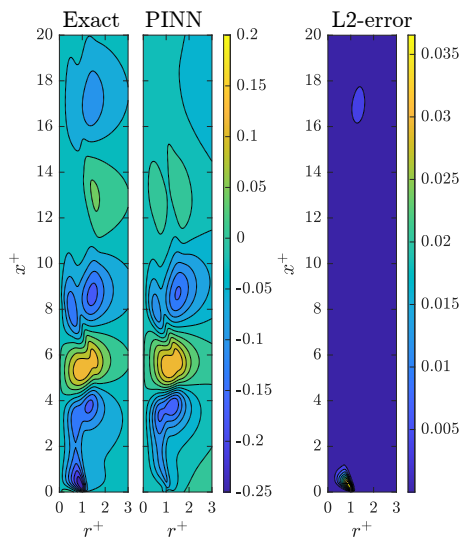}
    \caption{Comparison between the exact and reconstructed $v$ for the snapshot with maximal error ($t=0~\mathrm{s}$).}
    \label{fig:max_err_v}
\end{figure}

\revA{We infer that this higher error for that initial time instant is related to the absence of data (in time) that prevents an appropriate estimation of the time-derivative in the residuals as was observed in previous work \cite{Raissi2020}.} \revB{Here, it should be emphasised that the time between two snapshots was 1~ms (in contrast to the CFD timestep of $10^{-5}$ s) and, therefore, time-derivative quantities could not be inferred accurately from the measured fields, and even less so for the initial time instant.} \revA{In contrast to previous work \cite{Raissi2020}, in our case the error at the final time instants remains lower. This may be due to the fact that the PINN has appropriately recognised the periodic nature of the quantities to be reconstructed while for the initial time instant, this apparent periodicity had not yet been "detected" by the PINN.} However, despite the seemingly "large" error at the initial time (Fig.~\ref{fig:error_vel}), the actually reconstructed velocity field at that time instant, presented in Figs~\ref{fig:max_err_u} and \ref{fig:max_err_v} for the maximum error on $u$ and $v$ respectively, shows that the PINN still manages to reconstruct the main features of the velocity fields. The PINN appears to underestimates the axial velocity of the buoyant plume in the downstream region, while the radial velocity is mispredicted near the flame anchoring point.

\subsubsection*{Effect of noise}
\begin{figure}[!ht]
    \centering
    \includegraphics[width=67.7mm]{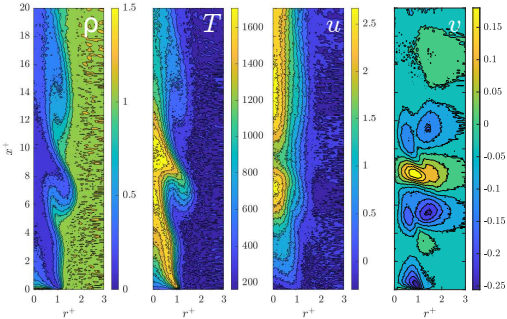}
    \caption{Typical noisy data fields with $\mathrm{SNR}=20~\mathrm{dB}$.}
    \label{fig:noisy_data}
\end{figure}

An additional test of the proposed reconstruction method is performed by considering the effect of noise on the reconstruction accuracy. For this analysis, the originally noise-free dataset is corrupted artificially by adding noise on all measured quantities ($\rho$, $p$, $T$) with a specified signal-to-noise ratio (SNR, based on each field average). Three levels of SNRs were considered: 20, 30 and 40~dB. Note that 20~dB refers to the highest noise level. A typical noisy dataset is shown in Fig.~\ref{fig:noisy_data} for $\mathrm{SNR}=20~\mathrm{dB}$. It can be seen that the data used for training by the PINN now exhibits noisy fluctuations. This makes the training of the PINN more complex as it has to identify how to smooth out the data to minimise the physics-based residual in Eq.~\eqref{eq:loss}. For each noisy dataset, a PINN with the exact same hyperparameters as for the noise-free case is trained following the same procedure as earlier.

\begin{figure}[!ht]
    \centering
    \includegraphics[width=67.7mm]{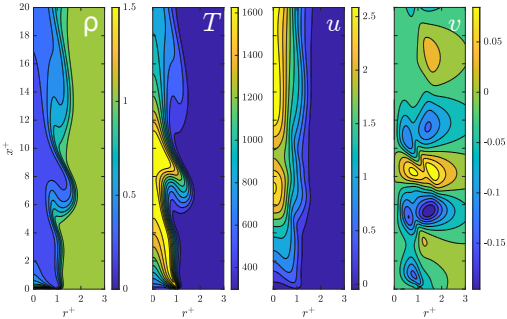}
    \caption{Inferred density, temperature and reconstructed velocity fields obtained from the PINN trained with noisy data with $\mathrm{SNR}=20~\mathrm{dB}$.}
    \label{fig:noisy_pred}
\end{figure}

A typical reconstruction obtained from the trained PINN is shown in Fig.~\ref{fig:noisy_pred} for the same time instant as the one shown in Fig.~\ref{fig:noisy_data}. It can be seen that, for the measured quantities ($\rho$ and $T$), the inferred fields from the PINN do not exhibit the noisy behaviour. This is achieved thanks to the physics-based loss in Eq.~\eqref{eq:loss}, which provides the means for the PINN to smooth out the inference of the measured quantities. Otherwise, fitting the data too closely, based uniquely on the mean-squared error, would lead to an excessively high physics-based loss due to the resulting spurious gradients. Instead, the trained PINN achieves a balance between fitting the noisy data and minimising the residual of the governing equations. This illustrates that the PINN can also act as a physics-based "denoiser" of measurement data. 

\begin{figure}[!ht]
    \centering
    \includegraphics[width=67.7mm]{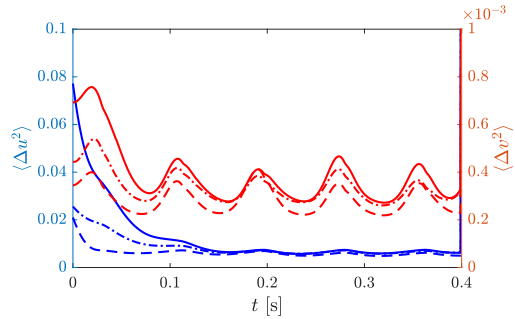}
    \caption{Evolution of MSE of the reconstruction of $u$ and $v$ when using noisy datasets. (full line: $\mathrm{SNR}=20~\mathrm{dB}$, dotted-dashed line: $30~\mathrm{dB}$, dashed line: $40~\mathrm{dB}$).}
    \label{fig:error_vel_noise}
\end{figure}

Regarding the reconstruction of the velocity field, it can be seen that the trained PINN reconstructed all features appropriately (compare velocity in Fig.~\ref{fig:noisy_pred} with Figs~\ref{fig:recons_velocity_u} and \ref{fig:recons_velocity_v}) indicating that the reconstruction from the PINN is robust with respect to noise. The resulting mean squared reconstruction error (computed with respect to the noise-free data) in function of time is shown in Fig.~\ref{fig:error_vel_noise} for all considered noise levels. Overall, the level of reconstruction accuracy is similar as compared to noiseless data (shown in Fig.~\ref{fig:error_vel}). In addition, as could be expected, the accuracy of the reconstruction decreases when larger noise levels are considered in the measured quantities. However, the overall accuracy remains within the range of what was obtained with the noiseless data indicating that the PINN is able to smooth out the measured data by using the equations and then use the relevant information to infer the appropriate reconstruction.

\subsubsection*{Effect of reduced measurements}

In this section, we analyse the reconstruction ability of the HFM when fewer measured quantities are considered. First, we consider the case when only the pressure and temperature measurements are available. The HFM framework is kept exactly the same as in the previous section and the only change is made by considering that measurements of density are unavailable. Therefore, the PINN has to reconstruct that quantity as well.
\revB{Initially this task was attempted without any assumption about the density, but the convergence of the PINN during training was relatively poor, since the reconstructed fields were not accurate enough. This was inferred to be due to the increased complexity of the optimisation problem solved by the PINN, where different combinations of reconstructed quantities ($\rho$, $u$ and $v$) could allow for a small loss during training.} 
\revB{To compensate for this, and to drive the PINN towards an appropriate solution,} during the pre-training phase (when the PINN is trained purely on data), a "physics-based" guess of the target density is introduced to drive the PINN towards predicting this guess (i.e. this physics-based guess replaces the density data during the pre-training). The physics-based guess for density is defined as:
\begin{equation}
    \label{eq:rho_guess}
    \rho_\mathrm{guess} = \frac{p_\mathrm{data}}{r_\mathrm{air} T_\mathrm{data}}
\end{equation}
where the subscript \emph{data} indicates that the value from the measured data is used and $r_\mathrm{air}=287.05~\mathrm{J/(kg\cdot K)}$ is the ideal gas constant of air, which is taken as a constant for the whole domain. This is evidently an approximation given the variations in composition (fuel and reaction products) occurring throughout the domain, but this allows to have an initial guess to drive the training of the PINN with this reduced set of information. 

\begin{figure}[!ht]
    \centering
    \includegraphics[width=67.7mm]{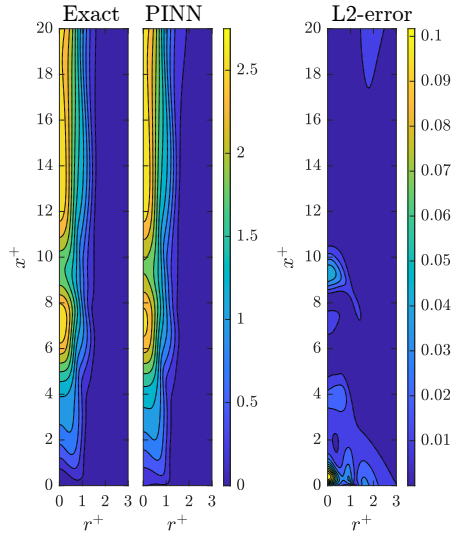}
    \caption{Comparison between actual and reconstructed $u$ velocity for the mid-time snapshot ($t=0.2~\mathrm{s}$). Superscript $+$ indicates coordinates normalised by $a$.}
    \label{fig:recons_2mes_u}
\end{figure}

\begin{figure}[!ht]
    \centering
    \includegraphics[width=67.7mm]{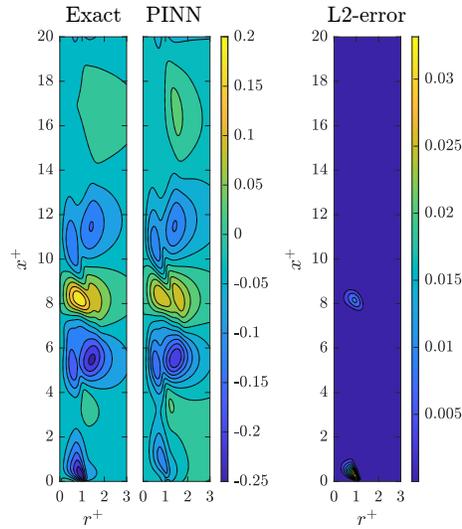}
    \caption{Comparison between actual and reconstructed $v$ velocity for the mid-time snapshot ($t=0.2~\mathrm{s}$). Superscript $+$ indicates coordinates normalised by $a$.}
    \label{fig:recons_2mes_v}
\end{figure}

Following this pre-training, the PINN is then further trained by using the physical equations as described in Eq.~\eqref{eq:loss}. The resulting reconstruction of the velocity fields when using such an approach is shown in Figs~\ref{fig:recons_2mes_u} and \ref{fig:recons_2mes_v} for the $u$ and $v$ velocity respectively at a representative time instance. It can be seen that the reconstruction accuracy is very similar to the case with three measurements shown previously. There is, however, a larger error locally close to the flame base at the fuel interface. 

\begin{figure}[!ht]
    \centering
    \includegraphics[width=67.7mm]{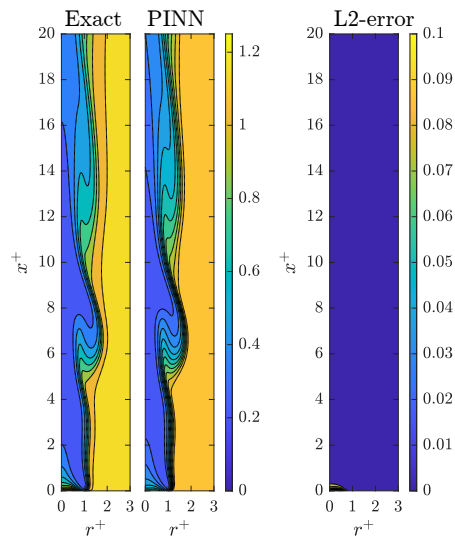}
    \caption{Comparison between actual and reconstructed density $\rho$ for the mid-time snapshot ($t=0.2~\mathrm{s}$). Superscript $+$ indicates coordinates normalised by $a$.}
    \label{fig:recons_2mes_rho}
\end{figure}

This latter error can be better understood from Fig.~\ref{fig:recons_2mes_rho}, where the reconstructed density field for the same time instant is shown. It can be seen that, overall, the density is very accurately reconstructed except near the fuel inlet. This error originates from the approximation made when proposing an initial guess for the density based on Eq.~\eqref{eq:rho_guess}. \revA{Since the ideal gas constant of stoichiometric heptane-air reaction products, $\approx 291~\mathrm{J/(kg\cdot K)}$, is very close to the ideal gas constant of air, this approximation is relatively accurate in all fuel-lean regions. 
In contrast, in the fuel-rich region near the fuel inlet}, this assumption is most inaccurate, therefore, leading to an inaccurate estimations of the density, which could not be fully compensated by the physics-based loss.

\begin{figure}[!ht]
    \centering
    \includegraphics[width=67.7mm]{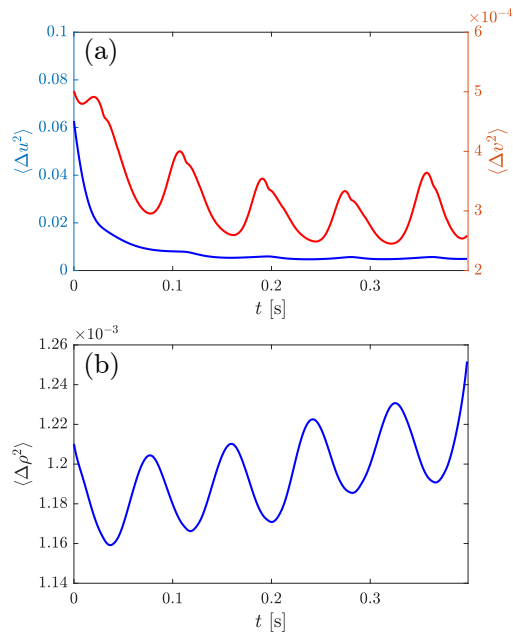}
    \caption{Evolution of MSE of the reconstruction of (a) $u$ and $v$ (b) $\rho$.}
    \label{fig:error_2mes}
\end{figure}

Nonetheless, despite this slightly inaccurate density reconstruction, the overall reconstruction accuracy of this two-measurement based PINN remains of the same order of magnitude as for the three-measurement based PINN in the previous section, as can be seen from the time-evolution of the average velocity reconstruction error in Fig.~\ref{fig:error_2mes}a, when compared to Fig.~\ref{fig:error_vel}. In both cases the error follows a similar evolution. In addition, Fig. \ref{fig:error_2mes}b shows the time-evolution of the density reconstruction error, where it can be seen that the error remains more or less constant with fluctuations related to the puffing frequency.

\begin{figure}[!ht]
    \centering
    \includegraphics[width=67.7mm]{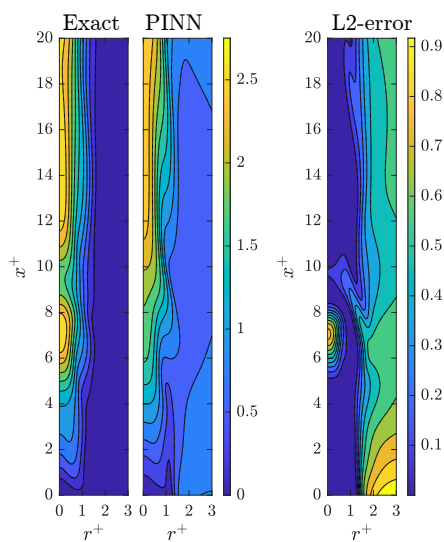}
    \caption{Comparison between actual and reconstructed $u$ velocity for the mid-time snapshot ($t=0.2~\mathrm{s}$). Superscript $+$ indicates coordinates normalised by $a$.}
    \label{fig:recons_1mes_u}
\end{figure}

\begin{figure}[!ht]
    \centering
    \includegraphics[width=67.7mm]{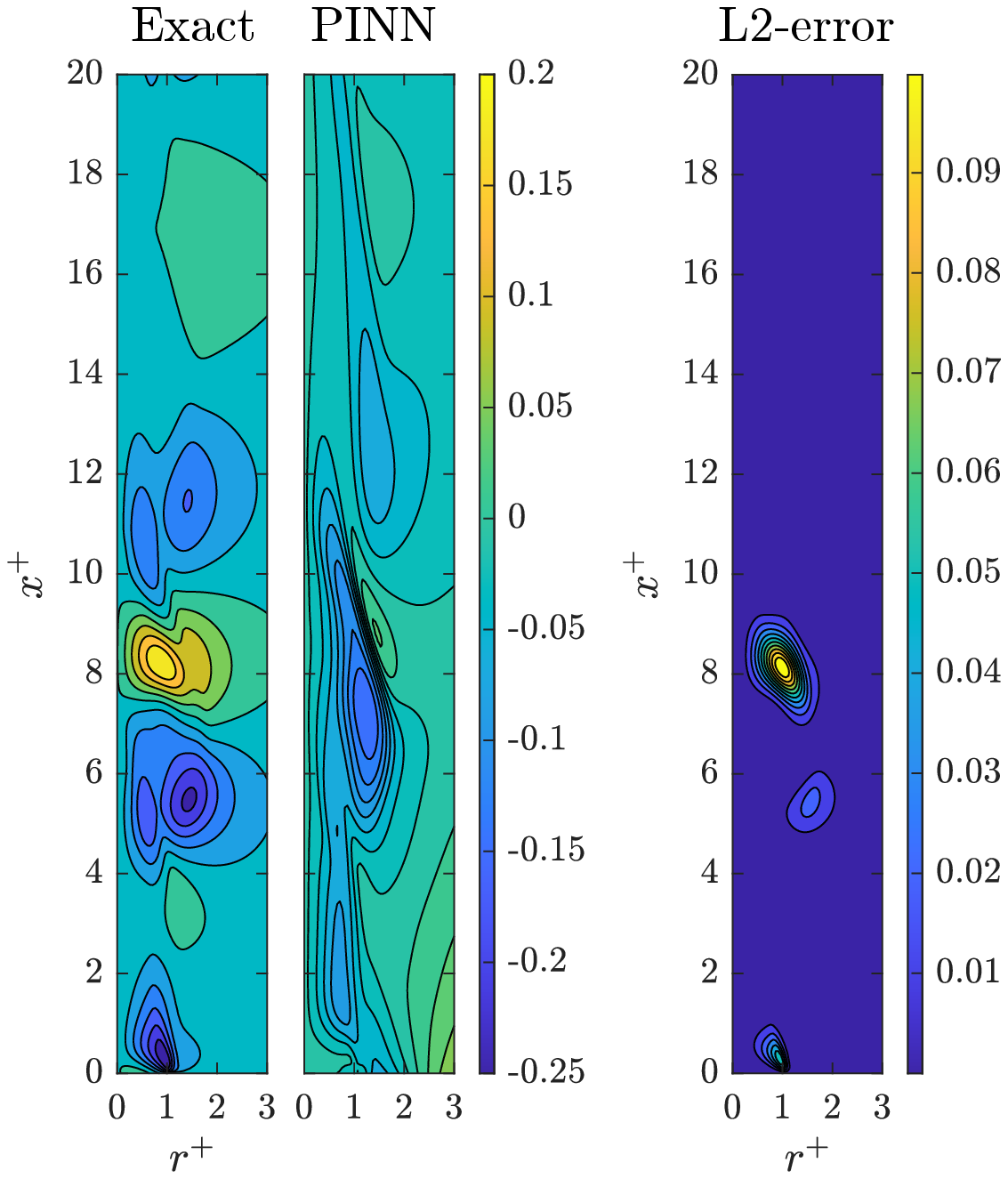}
    \caption{Comparison between actual and reconstructed $v$ velocity for the mid-time snapshot ($t=0.2~\mathrm{s}$). Superscript $+$ indicates coordinates normalised by $a$.}
    \label{fig:recons_1mes_v}
\end{figure}

\begin{figure}[!ht]
    \centering
    \includegraphics[width=67.7mm]{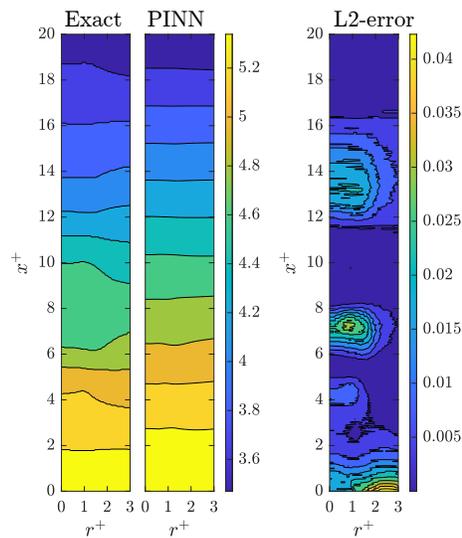}
    \caption{Comparison between actual and reconstructed pressure $p$ for the mid-time snapshot ($t=0.2~\mathrm{s}$). Superscript $+$ indicates coordinates normalised by $a$.}
    \label{fig:recons_1mes_p}
\end{figure}

Secondly, it is attempted to reconstruct the velocity, density and pressure fields solely from the measurements of temperature, to further analyse the reconstruction capability of the PINN. Here, temperature is chosen as it could be relatively straightforwardly measured in large-scale pool fires using infrared cameras, for example. Similarly as above, where a "guess" for the density was used for the pre-training of the PINN, a "guess" of the pressure field is also used to help during the pre-training of the PINN. The pressure field is assumed to follow a linear distribution with the streamwise coordinate, following:
\begin{equation}
    p_\mathrm{guess} = p_0 + \rho_\mathrm{air} g_x x
\end{equation}
where $\rho_\mathrm{air}$ is the density of air at standard conditions and $g_x = -9.81~\mathrm{m/s^2}$ is gravitational acceleration in axial direction. Then $\rho_\mathrm{guess}$ is estimated as in Eq.~\eqref{eq:rho_guess}, but using $p_\mathrm{guess}$ instead of $p_\mathrm{data}$. The values obtained in this manner for $\rho$ and $p$ are then used during the pre-training stage (i.e. they replace the data of density and pressure). The reconstructed velocity field (for the same time instant as in the previous section) is shown in Figs~ \ref{fig:recons_1mes_u} and \ref{fig:recons_1mes_v}. It can be directly observed that the reconstruction accuracy is lower than in the previous case specifically in the outer flame region (for $r^+>1.5$) and that the reconstructed $v$-velocity field (Fig.~\ref{fig:recons_1mes_v}) does not accurately reproduce all the features related to the puffing of the flame. This inaccuracy in the velocity reconstruction can be related to the inaccurate reconstructed pressure field (shown in Fig.~\ref{fig:recons_1mes_p}), where it can be seen that the radial variations of pressure are not accurately captured by the PINN. However, such variations are crucial as they induce the radial velocity. 

\begin{figure}[!ht]
    \centering
    \includegraphics[width=67.7mm]{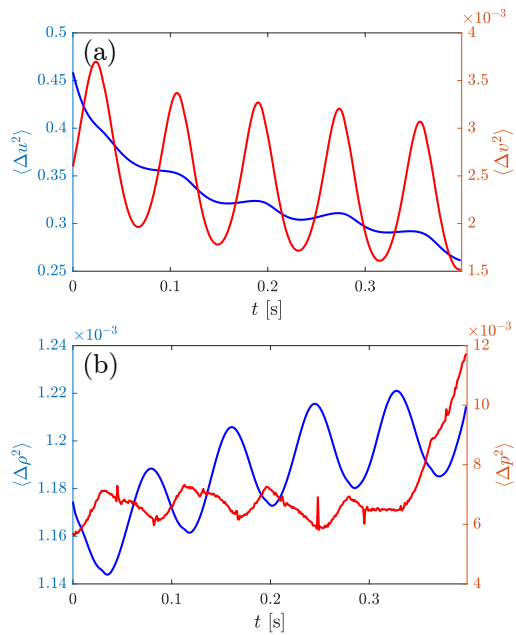}
    \caption{Evolution of MSE of the reconstruction of (a) $u$ and $v$ (b) $\rho$ and $p$.}
    \label{fig:error_1mes}
\end{figure}

Figure~\ref{fig:error_1mes} shows the evolution of the mean squared reconstruction error for the velocity field, $\rho$ and $p$. As expected, it can be seen that compared to the case where more measurements are available the error is larger (one order of magnitude larger for the $v$-velocity) indicating that relying solely on one measurement (here the temperature) to reconstruct all the other quantities of interest may be out of reach of the current PINN architecture. \revA{This result would suggests that we would generally need at least two measurements (of temperature and pressure) to be able to deduce an appropriate kinematic field in a reacting flow. It may be possible that another combination of measured quantities (such as temperature and one of the chemical species) could achieve a higher accuracy but such an analysis is left for future investigations.}

\revB{In addition, the observations above also highlight that, in the case when only one measurement becomes available, the optimisation problem that the PINN has to solve (i.e. the minimisation of the loss) becomes much more complex as there may exist various combinations of the other quantities to be reconstructed that could lead to a small loss values, but which are not the actual physical quantities. In other words, the PINN may reach a local minimum during its training which does not correspond to the actual flow solution. Such an observation shows the limit of the proposed PINN-based approach for the presented cased of a puffing pool fire.} 

The presently noted deficiencies in the predictions of velocity from temperature can be compared to relative success in a previous work by Cai et al.~\cite{Cai2021A102}, albeit limited to a qualitative validation. The most apparent difference between the cases lies in the presence of both large temperature variations ($\Delta T / T_0 > 1$) and significant variations of composition in the present pool fire, in contrast to small temperature variations, linearised density variations and negligible composition variations in the previous work. As a consequence, in the pool fire, larger errors are introduced when relating temperature with density and pressure. This problem is alleviated when two "measured" quantities are available. Evidently, two "measured" quantities allow to better describe two types of variations, temperature and composition. This may be related to the fact, that a diffusion flame structure cannot be described unambiguously by temperature alone. Nevertheless, variations of temperature and composition in a flame are not independent and could be related if the PINN was aware of the flame structure and able to detect the position of the flame through, for example, the measurements of a representative species.

\section{Conclusion} 
In this paper, we demonstrated for the first time the capability of the HFM framework to reconstruct unmeasured quantities from measured ones in the buoyancy-driven, reacting flow of a puffing pool fire. First, CFD simulations of pool fires at the onset of puffing were performed and validated against previous experiments to accurately reproduce the appropriate subcritical flame lengths, critical point and puffing frequency. The present simulations can, therefore, be seen as a surrogate for experimental data, but the simultaneous availability of all variable fields allows to explore the capabilities and limitations of HFM. The dataset obtained from CFD results was then leveraged to train the PINN and to show its reconstruction capabilities: inferring the velocity field from measurements of density, pressure and temperature -- but without any observations of the velocity field itself -- a high reconstruction accuracy was obtained. A further test was performed by considering noisy data. It was shown that the HFM framework could act as a physics-based "denoiser" as it could smooth out the added noise on the measured data. In addition, the PINN could still accurately reconstruct the main features of the velocity field, which shows that the proposed method is robust with respect to noisy data. In a final step it was attempted to reduce the number of "measured" fields required to predict the velocity field. Reasonable results were obtained when using density and pressure as inputs. However, it was not possible to make predictions from the temperature field alone. This emphasises the complexity of the present case involving both large variations of temperature and chemical composition. This may suggest that a PINN capable of predicting the behaviour of a flame needs to be able to identify the position and structure of the reaction zone.

This work demonstrates the potential of deep learning techniques for reconstructing unmeasured quantities in reacting flows in a physically consistent manner. This opens up possibilities for augmenting data obtained from experimental measurements in fire and combustion diagnostics.

In future work, the ability of this framework to reconstruct physical quantities in 3D and turbulent conditions will be explored. \revA{In addition, further analysis will be conducted on assessing which measurements are required to be able to infer a unique kinematic field}. The reconstruction of quantities from actual experimental data will also be investigated.


\section{Acknowledgements}

Photos of the puffing flame from experiments were kindly provided by W. Coenen from  Universidad Carlos III de Madrid. 


\bibliography{library,library_mps}


\end{document}